\documentclass[pre,aps]{revtex4}\begin{document}\title{Solidity of viscous liquids. IV. Density fluctuations}\author{Jeppe C. Dyre}\affiliation{DNRF centre  ``Glass and Time,'' Building 27 (IMFUFA),Roskilde University, Postbox 260, DK-4000 Roskilde, Denmark}\date{\today}\newcommand{\xik}{\xi_{\bf k}}\newcommand{\iomt}{i\omega\tau}\newcommand{\bu}{{\bf u}}\newcommand{\br}{{\bf r}}\newcommand{\bnul}{{\bf 0}}\newcommand{\bk}{{\bf k}}\newcommand{\bom}{{\bf\Omega}}\newcommand{\bnabla}{{\bf\nabla}}\newcommand{\thn}{\theta_0}\newcommand{\la}{\left\langle}\newcommand{\ra}{\right\rangle}\newcommand{\rk}[1]{{\rho_{\bf  {#1}}}}\newcommand{\fik}[1]{{\Phi_{\bf  {#1}}}(t)}\newcommand{\fnik}[1]{{\Phi^{(0)}_{\bf  {#1}}}(t)}\newcommand{\rok}{\rk\bk}\newcommand{\romk}{\rk{-\bk}}\newcommand{\gr}{\Gamma_{\rho}}\newcommand{\sti}{\tilde s}\newcommand{\tti}{\tilde t}\newcommand{\mt}{\langle\rho\rangle}\newcommand{\fnul}{\Phi}
\begin{abstract}
This paper is the fourth in a series exploring the physical consequences of the solidity of highly viscous liquids. It is argued that the two basic characteristics of a flow event (a jump between two energy minima in configuration space) are the local density change and the sum of all particle displacements. Based on this it is proposed that density fluctuations are described by a time-dependent Ginzburg-Landau equation with rates in k-space of the form $\Gamma_0+Dk^2$ with $D\gg\Gamma_0a^2$ where $a$ is the average intermolecular distance. The inequality expresses a long-wavelength dominance of the dynamics which implies that the Hamiltonian (free energy) may be taken to be ultra local. As an illustration of the theory the case with the simplest non-trivial Hamiltonian is solved to second order in the Gaussian approximation, where it predicts an asymmetric frequency dependence of the isothermal bulk modulus with Debye behavior at low frequencies and an $\omega^{-1/2}$ decay of the loss at high frequencies. Finally, a general formalism for the description of viscous liquid dynamics, which supplements the density dynamics by including stress fields, a potential energy field, and molecular orientational fields, is proposed.
\end{abstract}
\pacs{64.70.Pf}
\maketitle

\section{Introduction}

Glasses are made by cooling viscous liquids. The liquid relaxation time $\tau$ increases dramatically upon cooling, and the glass transition takes place when $\tau$ exceeds the inverse cooling rate: $\tau\gg 1/|d\ln T/dt|$. For cooling rates of order K/min one speaks about the calorimetric glass transition; glasses, however, may be produced by much higher cooling rates like the splat coolings of traditional metallic glasses, or much more slowly as when manufacturing large mirrors for astronomical telescopes by cooling gently over months. In all cases, by definition the glassy state has been arrived at when the liquid is no more in thermal equilibrium. The glassy state is not unique -- it depends on the thermal history of the system after it first fell out of equilibrium. In view of this inherent complexity of the glassy state it appears that a genuine understanding of a glass and its properties can come only from a proper understanding of the preceding much simpler equilibrium viscous liquid phase.

Are viscous liquids just {\it quantitatively} different from less-viscous liquids or are they, in fact, {\it qualitatively} different \cite{kau48,moo57,bue59,gol69,gol72,wyl79,bra84,dyr87}? On the one hand, one expects that all liquids obey the Navier-Stokes equation, albeit with a viscosity which in some cases is so large that it would take years to pour the liquid out of a cup; this suggests that the difference is just quantitative. On the other hand, consider the actual molecular motions in viscous liquids. Going back in time at least to Kauzmann's 1948 paper \cite{kau48}, these were always believed to be predominantly vibrational. The physical picture is that the molecules are caught in deep potential energy minima, vibrating billions and billions of times before rearranging to arrive at another potential energy minimum. Kauzmann described these rare {\it flow events} as ``jumps of molecular units of flow between different positions of equilibrium in the liquid's quasicrystalline lattice'' \cite{kau48}. This idea, which was elaborated upon in Goldstein's 1969 paper \cite{gol69}, was confirmed during the last decade in numerous computer simulations. In our interpretation the resulting physical picture \cite{note1} is:\begin{equation}\label{1}{\bf Viscous\,\, liquid\,\,\cong\,\,Solid\,\, that\,\, flows}\,.\end{equation}The implicit statement is: {\it Viscous liquid$\,\,\neq\,\,$ordinary liquid}. If the vibrations are averaged out, viscous liquid dynamics may be identified with the ``inherent dynamics'' consisting of jumps between the energy minima in the system's configuration space \cite{sch00}. Each minimum has a basin of attraction \cite{gol69,gol72}, known as an inherent structure. The concept of inherent structures was originally introduced by Stillinger and Weber as a way of thinking about liquids and solids in general \cite{sti83}, but the concept seems to be particularly useful for understanding the physics of highly viscous liquids. 

The liquid relaxation time $\tau$ is related to the (shear) viscosity $\eta$ and the instantaneous shear modulus $G_\infty$ by Maxwell's famous relation $\tau=\eta/G_\infty$ \cite{har76}. For ``ordinary'' less-viscous liquids like ambient water the viscosity is in the $10^{-3}$ Pa s range and the instantaneous shear modulus is typically of order ${\rm 10^9-10^{10}}$ Pa. Thus the relaxation time is barely one picosecond and comparable to a typical molecular vibration time. Viscous liquids approaching the calorimetric glass transition, on the other hand, have relaxation times of order 100-1000 seconds. These extremely long relaxation times reflect the fact that molecular motion has almost completely seized. The molecules still have the thermal velocity distribution, of course, but virtually all motion is vibrational like in a solid. Thus the incoherent (single-particle) diffusion constant $D_s$ is extremely small, and in the well-known expression for $D_s$ in terms of the velocity autocorrelation function $D_s=\int_0^\infty \langle v_x(0)v_x(t)\rangle dt$ there is a most delicate cancellation of contributions. This fact was emphasized in 1984 by Brawer \cite{bra84} who pointed out that if one wishes to apply conventional liquid state theory to viscous liquids approaching the calorimetric glass transition, the approximations used should be accurate to more than 15 digits in order to give reasonable results.

A further argument for viscous liquids being qualitatively different from the less-viscous liquids dealt with in traditional liquid state theory is the following. A liquid is characterized by a number of diffusion constants: The heat diffusion constant, the incoherent diffusion constant, the dynamic viscosity $\nu$ of the Navier-Stokes equation (i.e., the transverse momentum diffusion constant: viscosity/density), and the coherent diffusion constant (characterizing the long-wavelength decay of density fluctuations). For ``ordinary'' liquids these diffusion constants are all typically within one or two orders of magnitude of $10^{-7}  {\rm m^2/s}$. This is easy to understand from kinetic theory, according to which the diffusion constant is of order the mean-free path squared over the mean time between collisions of the diffusing entity. Rough estimates of these quantities are one Angstrom and 0.1 picosecond respectively, resulting in the value $10^{-7}$ ${\rm m^2/s}$. For a viscous liquid approaching the calorimetric glass transition, however, the approximate identity of diffusion constants breaks down: As the Maxwell relaxation time increases upon cooling from the less-viscous phase, the single-particle diffusion constant $D_s$ decreases roughly inversely proportionally to viscosity \cite{note2}. At the same time the dynamic viscosity $\nu$ increases. Thus the ratio $D_s/\nu$ changes from roughly $1$ to a number of order $10^{-30}$ just above the calorimetric glass transition. Such small dimensionless numbers are rare in condensed matter physics. Small numbers in physics usually signal a qualitative change and a simplification of some kind. It view of this it seems likely that the physics of highly viscous liquids are different from -- and somehow simpler than -- those of  ``ordinary'' liquids.

Viscous liquids have common features which are independent of the nature of the chemical bonds involved. These features \cite{ang00,rmp} may be summarized into three {\it non}'s: {\it Non-Debye} behavior of the main (alpha) relaxation process, {\it non-Arrhenius} viscosity (or relaxation time, $\tau\sim\eta$ because $\tau=\eta/G_\infty$), and {\it non-linearity} of relaxations following even relatively small temperature jumps. The last {\it non} is probably less basic because in most models {\it non-linearity} follows from the strong temperature dependence of $\tau$, but the {\it non-Debye} and {\it non-Arrhenius} behaviors constitute crucial and defining characteristics of viscous liquids. 

Given the universal features of viscous liquids, an obvious question is: Does the high viscosity -- translating into (\ref{1}) -- {\it in and of itself} make it possible to physically understand and mathematically derive the {\it non}'s? Since the high viscosity is caused by energy barriers large compared to $k_BT$, this seems to have been Goldstein's view when he in 1969 wrote: ``I am only conjecturing that whatever rigorous theory of kinetics we will someday have, processes limited by a high potential barrier will share some common simplifications of approach'' \cite{gol69}. 

The present paper is the fourth in a series (I-III) \cite{I,II,III} entitled {\it Solidity of viscous liquids} attempting to identify Goldstein's ``common simplifications of approach'' by extracting the physical consequences of Eq. (\ref{1}). The term solidity is meant as a concise way of referring to the ``solid-like-ness'' of viscous liquids. The first paper from our group utilizing solidity type arguments preceded the series. This was a joint publication with Olsen and Christensen from 1996 \cite{0} which proposed a model for the {\it non-Arrhenius} temperature dependence of $\tau$ (see also Ref. \cite{dyr98}). The idea is that the activation energy for a flow event is the work done in shoving aside the surroundings in order to -- in a brief moment -- create the extra space which by assumption is needed for the molecules to rearrange. According to this ``shoving'' model the activation energy is mainly shear elastic energy located in the surroundings of the rearranging molecules. Consequently, the activation energy is proportional to the instantaneous shear modulus, a quantity which is usually much more temperature dependent in viscous liquids than in crystals and glasses (or in ``ordinary'' liquids). The shoving model appears to work well for molecular liquids \cite{0}, but it is too early to tell whether the model is generally applicable \cite{rmp}. 

Paper (I) introduced the concept of a solidity length $l$ that determines the length scale below which a viscous liquid for all purposes behaves as a solid, albeit one that flows. In terms of the average intermolecular distance $a$ (defined by writing the volume per molecules as $a^3$), the alpha relaxation time $\tau$, and the high-frequency sound velocity $c$, the solidity length is given by \begin{equation}\label{2}l^4\,=\,a^3\,\tau\,c\,.\end{equation}This expression was derived by noting that a flow event is followed by the emission of a spherical sound wave; $l$ is determined by requiring that elastic equilibrium is just about established throughout a sphere with radius $l$ before the next flow event inside the sphere typically takes place \cite{I}. At the calorimetric glass transition the solidity length is several thousand Angstrom. 

In (II) we discussed anisotropic flow events and defined a parameter characterizing the anisotropy; it now appears, however, that this parameter is of only minor significance -- see below. The origin of the {\it non-Debye} alpha linear response functions was the subject of paper (III) where it was argued that a long-time-tail mechanism may explain what appears to be a generic property of the alpha process as observed, e.g., by dielectric relaxation \cite{ols01}: At high frequencies the loss follows an $\omega^{-1/2}$ decay once the effects of beta processes are minimized by going to sufficiently low temperatures (but still in the equilibrium liquid phase). It was further argued that the coherent diffusion constant $D$ is much larger than the incoherent diffusion constant $D_s$, implying that $D\gg a^2/\tau$. The latter  inequality, which is essential for the long-time-tail mechanism to work for a range of times {\it shorter} than $\tau$, i.e., above the alpha loss peak frequency, reflects a long-wavelength dominance of the dynamics. 

Long-time tails usually derive from a diffusion equation which implies decay rates $\propto k^2$ for the k-wavevector component of the relevant conserved variable \cite{pom75,kir02}. An explicit realization of the long-time-tail scheme of paper (III) was given subseqently in a simple model for dielectric relaxation \cite{dyr05}. This model is based on a conserved scalar field, the density, and a non-conserved vector field, the dipole density, with the simplest possible interaction term. When the model is solved in the Gaussian approximation to second order in the interaction strength between the two fields, one finds for the frequency-dependent dielectric constant \begin{equation}\label{2a}\epsilon(\omega)\,\propto\,  \frac{1}{\sqrt{1+\iomt}}+\frac{1}{\sqrt 2+\sqrt{1+\iomt}} + \frac{C}{1+i\omega\tau}\,.\end{equation}This expression reproduces the generic \cite{ols01} asymptotic behaviors of the alpha process: $-\epsilon''\propto\omega$ at low frequencies and $-\epsilon''\propto1/\sqrt\omega$ at high frequencies.

The present paper elaborates on the point made briefly in (III) and Ref. \cite{dyr05} that the density ``dispersion relation'' (the decay rate) realistically should include an additive constant and be of the form $\Gamma_0+Dk^2$ where $\Gamma_0=1/\tau$. We arrive at this result by first identifying the most important characteristics of a flow event. A new argument is given for the long-wavelength-dominance inequality $D\gg \Gamma_0a^2$. Then a model for density fluctuations is proposed and solved in the simplest approximation, predicting that the loss peak of the frequency-dependent isothermal bulk modulus is asymmetric towards the high-frequency side where the loss decays as $\propto\omega^{-1/2}$. Finally, we propose principles for the general description of viscous liquid dynamics.

\section{Basic characteristics of a flow event}
\subsection{A first approach}

The arguments of this paper refer to length scales below the solidity length $l$. For liquids approaching the calorimetric glass transition $l$ is so large that there are almost four orders of magnitude of length scales between $l$ and the intermolecular distance $a$. We argued above that in view of Eq. (\ref{1}) viscous liquid dynamics are basically to be identified to the ``inherent dynamics'' \cite{sch00} consisting of jumps between potential energy minima (inherent structures). This is what Goldstein envisaged in his emphasis on the importance of potential energy barriers much larger $k_BT$ \cite{gol69}. The fact that these barriers are large is directly responsible for the extremely large low-temperature viscosity, of course, but it also implies that there is a clear separation between the molecular vibrations and the much slower changes of the configurational degrees of freedom associated with jumps between the potential energy minima. In the resulting picture the vibrations are regarded as uncorrelated to the inherent dynamics, and only the latter contribute to autocorrelation-function variations on the time scale of the alpha relaxation. 

A flow event is a jump from one potential energy minimum to another. Leaving aside the interesting question why the alpha-process activation energy increases upon cooling, and whether this as in the shoving model \cite{0} reflects the fact that the high-frequency elastic constants increase upon cooling, we shall nevertheless use shoving-model type arguments below. This is done by comparing the situation {\it before} and {\it after} a flow event, and not as in the shoving model {\it before} and {\it at} the barrier maximum of the flow event. 

It seems compelling that flow events are localized in space; this is also what is observed in the numerous computer simulations that now have been published. Justified by Eq. (\ref{1}) we utilize arguments from the theory of solid elasticity \cite{lan70} by regarding the liquid as an isotropic solid in which the flow event takes place. As a simple model, suppose that the flow event is radially symmetric. If it takes place at $\br=\bf 0$, the radial displacement field induced in the surroundings may be written $\bu(\br)=u_r(r)\br/r$ where $r=|\br|$. To determine $u_r(r)$ we combine the equation of elastic equilibrium after the flow event -  zero divergence of the stress tensor: $\partial_i\sigma_{ik}=0$ -- with the stress-strain relation $\sigma_{ik}=Ku_{ll}\delta_{ik}+2G(u_{ik}-u_{ll}\delta_{ik}/3)$, where $K$ and $G$ are the bulk and shear moduli and $u_{ik}=(\partial_iu_k+\partial_ku_i)/2$ is the strain tensor. This leads to\begin{equation}\label{3}(K+\frac{G}{3})\bnabla(\bnabla\cdot\bu)\,+\,G\nabla^2\bu\,=\,{\bf 0}\,.\end{equation}Because the displacement is radial one has $\bnabla\times\bu={\bf 0}$ which, via the vector identity $\bnabla\times(\bnabla\times\bu)=\bnabla(\bnabla\cdot\bu)-\nabla^2\bu$, implies that $\bnabla(\bnabla\cdot\bu)=\nabla^2\bu$. When substituted into Eq. (\ref{3}) this leads to $\bnabla(\bnabla\cdot\bu)={\bf 0}$, or $\bnabla\cdot\bu=C_1$. The radial displacement is found by solving the equation$\bnabla\cdot\bu\equiv r^{-2}\partial_r(r^2u_r)=C_1$, leading to $u_r=C_2r^{-2}+C_1 r/3$. Since the latter term diverges as $r\rightarrow\infty$, we must have $C_1=0$. In conclusion, the displacement field is analogous to the Coulomb electric field of a point charge: \begin{equation}\label{4}\bu\,\propto\, \frac{\br}{r^3}\,.\end{equation}The relative density change is $-\bnabla\cdot\bu$ \cite{lan70} which is zero except at $\br=\bf 0$. Thus in this macroscopic and radially symmetric description there is no density change in the surroundings of the flow event. A decrease of density at the flow event center induces a positive radial expansion which results in the same particle flux through all spheres centred at the flow event. This means that for any spherical shell surrounding the flow event, thick or thin, just as many molecules enter the shell from the inside as leave it on the outside. In effect, a flow event corresponds to a three-dimensional version of Hilbert's hotel, the infinite hotel which -- even when totally occupied -- makes room for an extra guest by asking all guests to move to one higher room number. 

The above analysis is oversimplified because of the assumption of spherical symmetry. Nevertheless, the result that the displacement field far from the flow event center varies as $1/r^2$ is correct and general. To see this, note that the effect of one flow event on its surroundings may be mimicked by first imagining a tiny sphere surrounding the flow event center. If the molecules inside the sphere are removed, the effect of the change of positions of the molecules before and after the flow event may be reproduced by external forces acting suitably on the surface of the sphere -- forces that must sum to zero. An external force acting on a point in an elastic medium introduces a momentum flux spreading to infinity; since the stress tensor $\sigma$ is the momentum flux density, we conclude that $\sigma\propto 1/r^2$ as $r\rightarrow\infty$. The stress tensor is given by first order derivatives of the displacement vector field; thus one expects that $|\bu|\propto 1/r$. For forces summing to zero, however, we get $|\bu|\propto 1/r^2$, just as the potential from an electric dipole varies as $1/r^2$ whereas it varies as $1/r$ from a point charge. As shown elsewhere \cite{condmat}, even in the most anisotropic case more than 90\% of the elastic energy in the far-field surroundings is shear elastic energy, so the assumption that there are no density changes in the surroundings is a good approximation for describing the long-ranged effects of a flow event.

Flow events may be regarded as instantaneous on length scales below the solidity length. Numbering the flow events consecutively after the time they take place, $t_\mu$, if $\br_\mu$ is the center of the $\mu$'th flow event and the number $b_\mu$ measures its magnitude, the above considerations translate into the following dynamic equation for the density $\rho(\br,t)$ in a coarse-grained description:\begin{equation}\label{5}\dot\rho(\br,t)\,=\,\sum_\mu b_\mu\delta(\br-\br_\mu)\delta(t-t_\mu)\,.\end{equation}Equation (\ref{5}) is not inconsistent with particle conservation, of course. Nevertheless, a field obeying Eq. (\ref{5}) does have the appearance of not being conserved, because density changes at one point in space do not affect the density elsewhere. A field of non-interacting spins fluctuating randomly in time  would be described by a similar time evolution equation, and there is nothing conserved by such a spin field. To summarize, the ``Hilbert's hotel effect'' deriving from solidity is not inconsistent with the fact that the particle number is obviously and trivially conserved, but it implies that density {\it acts as if} it is a non-conserved field. Although we show below that Eq. (\ref{5}) is too simple to reflect all relevant features of density fluctuations, this conclusion remains valid.The interpretation of the flow event magnitude $b_\mu$ is found by integrating Eq. (\ref{5}) over a brief period of time including only the flow event taking place at $t_\mu$. This leads to a density change equal to $b_\mu\delta(\br-\br_\mu)$, so the number of particles inside any volume which includes the flow event center $\br_\mu$ changes by precisely $b_\mu$. Note that via the continuity equation $\dot\rho+\bnabla\cdot{\bf J}=0$ and the identity $\bnabla\cdot(\br/r^3)=4\pi\delta(\br)$, Eq. (\ref{5}) corresponds to the following expression for the particle current density in the coarse-grained description\begin{equation}\label{6} {\bf J}(\br,t)\,=\,-\sum_\mu \frac{b_\mu}{4\pi}\,\frac{\br-\br_\mu}{|\br-\br_\mu|^3}\,\delta(t-t_\mu)\,.\end{equation}

\subsection{A more detailed treatment}

To study flow events in more detail we consider the induced density changes by going to k-space. If the liquid consists of $N$ molecules in volume $V$ each with position $\br_j$, the variable $\rok$ defined by\begin{equation}\label{7}\rok\,=\,\frac{1}{\sqrt N}\sum_j e^{i\bk\cdot\br_j}\,.\end{equation}Normalizing in this way is convenient because it makes the fluctuations independent of sample size in the $V\rightarrow\infty$ limit where the static structure factor is given by $S(k)=\langle|\rok|^2\rangle$  \cite{boonyip}.

As usual, the k-vectors are restricted to values compatible with periodic boundary conditions, i.e., having an integer number of periods in the volume $V$ in all three axis directions. We are particularly interested in small k-vectors. If the molecular displacements induced by a single flow event are denoted by $\Delta\br_j$, by a first order Taylor expansion the change of $\rk\bk$ for small k is given by\begin{equation}\label{8}\delta\rok\,=\,\frac{1}{\sqrt N}\sum_j e^{i\bk\cdot\br_j}i\bk\cdot\Delta\br_j\,.\end{equation}Since particle displacement are unlikely to be much larger than the intermolecular distance $a$, Eq. (\ref{8}) applies whenever $ka\ll 1$. As discussed in (III) and Ref. \cite{dyr06}, for realistic viscous liquid samples momentum is not conserved because the transverse momentum diffusion constant (the kinematic viscosity $\nu=\eta/\langle\rho\rangle$) is so large that on the time scale of the alpha relaxation momentum is unavoidably exchanged between the sample holder and the liquid. Thus momentum conservation is irrelevant for viscous liquid dynamics just as it is, e.g., for the description of point defect diffusion in crystals, and the sum of all particle displacements induced by a flow event $\Delta{\bf R}=\sum_j\Delta\br_j$ is generally non-zero. In our present view the vector $\Delta{\bf R}$ is a more important measure of flow event anisotropy than the ``quadrupolar'' parameter introduced in (II), where it was implicitly assumed that $\Delta{\bf R}=\bf 0$.

If the $\mu$'th flow event is centered at $\bf r_\mu$, because the largest displacements take place close to $\bf r_\mu$, for small k it is tempting to argue (III) that it is a good approximation to replace all exponentials by $\exp(i\bk\cdot\br_\mu)$, leading to\begin{equation}\label{9}\delta\rok\,\cong\,\frac{e^{i\bk\cdot\br_\mu}}{\sqrt N}i\bk\cdot\Delta{\bf R}\,.\end{equation}This implies that $|\delta\rok|^2\propto k^2$ for small k. Equation (\ref{5}), on the other hand, which we argued provides a good coarse-grained description, implies that $|\delta\rok|^2$ is constant for small k. Which is right? In fact, both are partially correct as we now proceed to show.

The starting point is Eq. (\ref{8}) (still assuming small k). The calculation leading to Eq. (\ref{9}) is not quite correct, however. This is because, although displacements far from the flow event are small ($\propto 1/r^2$), there are many molecules far away ($\propto r^2$) and their contributions cannot be ignored. A more detailed analysis proceeds as follows. Suppose a flow event of magnitude $b$ is located at $\br={\bf 0}$. As a reasonable first approximation we use Eq. (\ref{6}) which implies that the total particle flux due to this flow event through the perpendicular area $dA$ at the distance $r$ from ${\bf 0}$ is equal to $(-b/4\pi)(dA/r^2)$. Identifying this flux with $\mt udA$, where $u$ is the particle displacement, leads to $u=-(b/4\pi r^2\mt)$. Thus the displacement of the $j$'th molecule, $\Delta\br_j$, is given by $\Delta\br_j=-(b/4\pi\mt) (\br_j/r_j^3)$. In terms of the density $\rho(\br)\equiv\sum_j\delta(\br-\br_j)$ Eq. (\ref{8}) thus becomes\begin{equation}\label{10}\delta\rok\,=\,\frac{-b}{4\pi\sqrt N}\int_V \frac{\rho(\br)}{\mt} \frac{i\bk\cdot\br}{r^3}e^{i\bk\cdot\br}\,d\br\,.\end{equation}If the flow event magnitude $b$ is uncorrelated to other quantities, the ensemble average of the absolute square is given by\begin{equation}\label{11}\langle|\delta\rok|^2\rangle\,=\,\frac{\langle b^2\rangle}{16\pi^2 N}\int_V\frac{\langle\rho(\br) \rho(\br')\rangle}{\mt^2}\, \frac{\bk\cdot\br}{r^3}\frac{\bk\cdot\br'}{r'^3}e^{i\bk\cdot(\br-\br')}d\br d\br'\,.\end{equation}At first sight this expression appears to confirm that $\langle|\delta\rok|^2\rangle\propto k^2$, but that is not correct: As shown in the Appendix, because the term $\langle\rho(\br) \rho(\br')\rangle$ for $|\br-\br'|\rightarrow\infty$ becomes constant, one has  \begin{equation}\label{12}\langle|\delta\rok|^2\rangle\,=\, \frac{\langle b^2\rangle}{N}\,\Big[1\,+\,\frac{1}{N}\sum_{\bk'} S(k') \left(\frac{\bk\cdot(\bk+\bk')}{(\bk+\bk')^2}\right)^2\,\Big]\,.\end{equation}For small k this implies that\begin{equation}\label{12a}\langle|\delta\rok|^2\rangle\,\propto\, 1+C(ka)^2\,,\end{equation}where $C\sim(1/N)\sum_{\bk'}S(k')/[(k'a)^2]\sim 1$.

\section{Density fluctuations}
\subsection{General framework}

Based on the above considerations we now seek a model for density fluctuations in equilibrium viscous liquids. To simplify matters it is assumed that density is the only relevant variable, although eventually other variables should definitely be included in the description (Sec. IV). The basic assumption is that the density obeys a standard Langevin equation \cite{van86}. For discrete degrees of freedom $Q_1,..., Q_n$ the Langevin equation starts from a Hamiltonian $H(Q_1,...Q_n)$, and the dynamics are given by the equations $\dot Q_i= -\Gamma_i \partial (\beta H)/\partial Q_i+\xi_i(t)$ where $\beta=1/k_BT$ and $\xi_i(t)$ is a Gaussian white noise term obeying $\langle\xi_i(t)\xi_j(t')\rangle=2\Gamma_i\delta_{ij}\delta(t-t')$. These equations give the correct canonical equilibrium probability for the average occupation in configuration space \cite{van86}.

In the present case the Hamiltonian is the free energy written as a functional of the density field. In terms of the complex density field variables $\rok$ the Langevin equation looks as follows with a complex noise term:\begin{equation}\label{13}\dot\rho_{\bf k}\,=\,-\Gamma_k \frac{\partial(\beta H)}{\partial \romk}\,+\,\xik(t)\,.\end{equation}The noise term obeys $\xik^*(t)=\xi_{-\bf k}(t)$ and $\langle\xik(t)\xi_{\bf k'}^*(t')\rangle=2\Gamma_k\delta_{\bf k,k'}\delta(t-t')$; because $\rho_{\bf k}^*=\rho_{-\bf k}$ Eq. (\ref{13}) is equivalent to two independent real Langevin equations, one for the real part of $\rho_{\bf k}$ and one for its imaginary part.  To determine the k-dependence of the rate $\Gamma_k$ we note that when Eq. (\ref{13}) is integrated over a short time interval, the noise term dominates. Thus if the short time interval is $\Delta t$, the magnitude of the change $\Delta\rok$ is given by $\langle|\Delta\rok|^2\rangle=2\Gamma_k\Delta t$. On the other hand, because flow events are uncorrelated over short time spans, from Eq. (\ref{12a}) one finds that $\langle|\Delta\rok|^2\rangle\propto\Delta t (1+C(ka)^2)$ for small k. In conclusion, for small k the rate $\Gamma_k$ is of the form \cite{rho_note}\begin{equation}\label{14}\Gamma_k\,=\,\Gamma_0+Dk^2\,.\end{equation}We shall assume that this expression applies for all $k$. A conserved field is characterized by $\Gamma_k\propto k^2$ for $k\rightarrow 0$ \cite{hoh77}, so Eq. (\ref{14}) expresses the fact that density has the appearence of a non-conserved field. According to the calculation of last section $C\sim 1$, which implies that $D\sim\Gamma_0a^2$. This calculation, however, assumes much more symmetry than realistically may be expected. Thus Eq. (\ref{4}) is based on the macroscopic elasticity theory describing an isotropic and homogeneous solid. Surely a viscous liquid is solid like, but it is neither homogeneous nor isotropic on the molecular scale, and although the fact that the displacements far from a flow event vary as $1/r^2$ remains valid, one wouldn't expect Eq. (\ref{4}) to be accurate. Violations of isotropy and homogeneity easily lead to a much larger flow event induced $\Delta\bf R$ than found in the calculations leading to Eq. (\ref{12a}) (similarly, local correlations of flow events as seen, e.g., in the observations of strings of flow events in some computer simulations \cite{sch00,strings}, also severely violate isotropy). In conclusion, since the $C$ term of Eq. (\ref{12a}) is determined by the magnitude of $\Delta\bf R$ which is a measure of the flow event anisotropy, it appears likely that $C\gg 1$, or equivalently  \begin{equation}\label{14a}D\,\gg\, \Gamma_0\, a^2\,.\end{equation}We shall henceforth assume that this inequality is obeyed. In (III) Eq. (\ref{14a}) was justified by arguing that $D$ is much larger than the incoherent diffusion constant. In neither case, however, has Eq. (\ref{14a}) been rigorously proved and it remains an assumption which is justified by physical arguments. 

In principle, all k-vectors consistent with periodic boundary conditions are allowed, but since there are just a finite number of molecules, one should only include the $N$ smallest k-vectors -- for larger k-vectors the $\rk{\bk}$'s become redundant. This means that there is an implicit cut-off in k-space, $k_c$, which is easily shown to be given by $k_ca\sim 1$. In view of this, Eq. (\ref{14a}) implies that there is a range of allowed k-vectors where the $Dk^2$ term of Eq. (\ref{14}) dominates over the $\Gamma_0$ term. As shown below and in Ref. \cite{dyr05} this makes it possible to understand the generic $\omega^{-1/2}$ high-frequency decay of the alpha loss as a consequence of a long-time tail mechanism operating over a range of times short compared to the alpha relaxation time.

The inequality Eq. (\ref{14a}) expresses a {\it long-wavelength dominance} of the dynamics. This assumption, which {\it a posteriori} justifies our focus on small k vectors, makes it possible to simplify the Hamiltonian considerably. A number of scattering experiments have looked for diverging length scales as the glass transition is approached, but found none. The consensus is that viscous liquids have no long-ranged static (i.e., equal time) density correlations. Thus, if the dynamics are dominated by the long-wavelength behavior, a model with no equal-time spatial density correlations should suffice. In field-theory terms this means that the Hamiltonian may be assumed to be ``ultra local,'' i.e., without the usual gradient term $(\nabla\rho)^2$ or other terms coupling fields at different points in space. Although this is rather unusual from a general field-theory perspective, note that a simple example of an ultra-local field theory is the free energy functional for an ideal gas. More generally, the Ramakrishnan-Yussouff density functional \cite{ram79} becomes ultra local in descriptions which are coarse-grained on length scales beyond the correlation length of the direct correlation function.

Thus we assume that the dimensionless Hamiltonian functional, when scaled by the inverse temperature $\beta$, is of the form\begin{equation}\label{14b}\beta H\ =\ \mt\int_Vd\br \left[\frac{1}{2A}\left(\frac{\rho(\br)-\mt}{\mt}\right)^2\,+\,\frac{\lambda}{3}\left(\frac{\rho(\br)-\mt}{\mt}\right)^3\,+\, ...\right]\,.\end{equation}The third order term is crucial for the results derived below; to ensure stability, however, there must be further higher order even terms. To transform this expression into k-space, we first note that Eq. (\ref{7}) implies that $\rho(\br)=(\sqrt N/V)\sum_\bk \rk\bk\exp(-i\bk\cdot\br)$, which in turn implies that $(\rho(\br)-\mt)/\mt=\sum_{\bk\neq 0}\rk\bk\exp(-i\bk\cdot\br)/\sqrt N$. When this is substituted into Eq. (\ref{14b}), the dimensionless Hamiltonian becomes\begin{equation}\label{15}\beta H\, =\,\frac{1}{2A}\sum_{\bk} \rk\bk \rk {-\bk}\,+\,\frac{\lambda}{3\sqrt N}\sum_{\bk,\bk'}\rk\bk \rk {\bk'} \rk{-\bk-\bk'}\,+\,...\,,\end{equation}where it is implicitly understood that no terms with $\bk={\bf 0}$ appear (because a system with fixed volume is considered, for $\bk={\bf 0}$ $\rk{\bk}$ is not a dynamic degree of freedom).

Once the dynamics have been specified one can calculate the density autocorrelation function as a function of time, the Laplace transform of which gives the dynamic structure factor. To the best of the author's knowledge, there are no data for this quantity for highly viscous liquids. But by the fluctuation-dissipation theorem the $k\rightarrow 0$ limit of the density autocorrelation function determines the macroscopic frequency-dependent isothermal bulk modulus. It is convenient to introduce the notation\begin{equation}\label{16}\fik{k}\,=\,\langle\rk{k}(0)\rk{-k}(t)\rangle\,.\end{equation}Obviously, $\fik{k}=\fik{-k}$ by time-reversal and parity invariance. It is possible to establish an exact equation for $\fik{k}$ by use of the following general theorem \cite{III,dyr05}: If $Q_i$ obeys a Langevin equation of the form (no sums over $i$) $\dot Q_i=-\Gamma_i\partial_i (\beta H) + \xi_i(t)$, one has $d^2/dt^2\la Q_i(0)Q_i (t)\ra=\Gamma_i^2\la \partial_i (\beta H)(0)\partial_i (\beta H)(t)\ra$. Substituting Eq. (\ref{15}) into this identity leads to\begin{equation}\label{17}\frac{d^2}{dt^2}\fik{k}\,=\,\Gamma_k^2\left\langle\Big(\frac{\rk\bk(0)}{A}+\frac{\lambda}{\sqrt N}\sum_{\bk'}\rk{\bk+\bk'}(0)\rk{-\bk'}(0)+...\Big)\Big(\frac{\rk{-\bk}(t)}{A}+\frac{\lambda}{\sqrt N}\sum_{\bk''}\rk{-\bk-\bk''}(t)\rk{\bk''}(t)+...\Big)\right\rangle\,.\end{equation}Approximations are needed in order to proceed. The simplest approximation is the Gaussian approximation which leads to non-linear self-consistent equations.

\subsection{The Gaussian approximation}

The simplest nontrivial case is when one ignores the higher-order terms of Eq. (\ref{15}), a procedure which is justified when these terms are so small that they do not significantly influence the density autocorrrelation function. Using the well-known fact that for variables with zero mean distributed according to a Gaussian one has $\langle x_1x_2x_3x_4\rangle=\langle x_1x_2\rangle\langle x_3x_4\rangle+\langle x_1x_3\rangle\langle x_2x_4\rangle+\langle x_1x_4\rangle\langle x_2x_3\rangle$ whereas averages of odd order are zero, Eq. (\ref{17}) in the Gaussian approximation becomes\begin{equation}\label{18}\frac{d^2}{dt^2}\Phi_{\bk}(t)\,=\,\Gamma_k^2\left( \frac{\fik{k}}{A^2}\,+\,2\frac{\lambda^2}{N}\sum_{\bk'}\fik{\bk+\bk'}\fik{-\bk'}\,+\, ...\right)\,.\end{equation}We keep only the terms relevant for the below calculation; as mentioned the inclusion of, e.g., a small fourth order term in the Hamiltonian may be ignored for calculating the density autocorrelation function -- its inclusion would simply lead to a minor renomalization of $A$ and a small $\Phi^3$-type term.

The $k\rightarrow 0$ limit of $\fik\bk$ may be determined analytically to second order in $\lambda$ by proceding as follows. To zeroth order in $\lambda$ the Hamiltionan is quadratic, implying for the equal time average that $\langle\rk\bk\rk{-\bk}\rangle=A$. Thus to zeroth order one has $\fnik\bk=A\exp(-\Gamma_\bk t/A)$, which is substituted into the perturbing term of Eq. (\ref{18}). Because of Eq. (\ref{14a}) the k-sum may be evaluated by extending the k-integration to infinity (recall that $\mt=a^{-3}$):\begin{eqnarray}\label{19}2\frac{\lambda^2}{N}\sum_{\bk'}\fnik{\bk'}\fnik{-\bk'}\,&=&\,2\frac{\lambda^2A^2}{N}\int V\frac{d\bk'}{(2\pi)^3}\,e^{-2(\Gamma_0+Dk'^2)t/A}\nonumber\\\,&=&\,\frac{2\lambda^2A^2}{(2\pi)^3\mt}e^{-2\Gamma_0t/A}\int_0^\infty dk'\,4\pi k'^2\,e^{-2Dk'^2t/A}\nonumber\\\,&=&\,\frac{2\lambda^2A^2}{(2\pi)^3\mt}e^{-2\Gamma_0t/A}4\pi\left(\frac{2Dt}{A}\right)^{-3/2}\frac{\sqrt\pi}{4}\\\,&=&\,\frac{\lambda^2A^{7/2}}{8\sqrt 2 (\pi D)^{3/2}\mt}t^{-3/2}e^{-2\Gamma_0t/A}\nonumber\,.\end{eqnarray}If the $k\rightarrow 0$ limit of $\fik\bk$ is denoted by $\fnul(t)$, we thus arrive at the following equation for $\fnul(t)$ in terms of the dimensionless time $\tilde t\equiv \Gamma_0t/A$\begin{equation}\label{20}\ddot\fnul(\tti)\,=\,\fnul(\tti)\,+\,\Lambda \tti^{-3/2}e^{-2\tti}\,,\end{equation}where $\Lambda=\lambda^2A^4(\Gamma_0/\pi D)^{3/2}/(8\sqrt 2\mt)$. The general solution of this differential equation obeying $\Phi(\tti\rightarrow\infty)=0$ is \begin{equation}\label{21}\Phi(\tti)\,=\,\Lambda\int_{\tti}^\infty \sinh({\tti}'-{\tti})e^{-2\tti'}\tti'^{\rm-3/2}d\tti'\,+\,C\,e^{-\tti}\,.\end{equation}This result applies to second order in $\lambda$ and assumes that the first term is small, but at this stage it is not possible to realistically estimate the relative weights of the two terms.

According to the fluctuation-dissipation theorem, if $v$ is a large subvolume of $V$, $\Delta v(t)\equiv v(t)-v(0)$, and $L(s)$ is the Laplace transform of $\langle\Delta v^2(t)\rangle$ evaluated at the complex variable $s\equiv i\omega$, the frequency-dependent isothermal bulk modulus $K_T(\omega)$ is given by \begin{equation}\label{22}K_T(\omega)\,\propto\,\frac{1}{s L(s)}\,.\end{equation}Besides the relaxational density fluctuations described by Eq. (\ref{13}) there are always the fast vibrational density fluctuations. In any reasonable model these two types of fluctuations are uncorrelated. Consequently, if $\Delta v_r(t)$ is the relaxational volume change over time $t$ and $\Delta v_v(t)$ the vibrational analogue,  one has $\langle\Delta v^2(t)\rangle=\langle\Delta v_r^2(t)\rangle+\langle\Delta v_v^2(t)\rangle$. On time scales much longer than phonon times the vibrational volume mean-square displacement is independent of time and one may write $\langle\Delta v^2(t)\rangle=C+\langle\Delta v_r^2(t)\rangle$. Thus $L(s)$ becomes $C/s$ plus the Laplace transform of $\langle\Delta v_r^2(t)\rangle$, a quantity that is proportional to $\Phi(0)-\Phi(t)$. Equation (\ref{21}) implies that $\Phi(0)-\Phi(t)\propto \sqrt t$ at short times (i.e., whenever $\tti\ll 1$ or equivalently $t\ll \tau$ where $\tau\equiv A/\Gamma_0$), whereas for $t\gg \tau$ $\Phi(t)$ goes exponentially to zero. These asymptotic behaviors imply that $K_T(\omega)\propto 1-C_1/\sqrt{i\omega\tau}$ for $\omega\tau\gg 1$ and $K_T(\omega)\propto 1+C_2(i\omega\tau)$ for $\omega\tau\ll 1$. For the imaginary part of the frequency-dependent bulk modulus (the loss) one finds $K_T''(\omega)\propto \omega^{-1/2}$ for $\omega\tau\gg 1$ and $K_T''(\omega)\propto \omega$ for $\omega\tau\ll 1$. The model thus predicts bulk modulus loss peaks which are asymmetric towards the high-frequency side. This is what is always observed for the dielectric and shear mechanical loss peaks, but unfortunately there are only few measurements of the frequency-dependent adiabatic bulk modulus (and none of the isothermal frequency-dependent bulk modulus). The only published adiabatic measurement known to this author (on glycerol \cite{chr94}) is not inconsistent with these predictions -- it is described by a stretched exponential relaxation function with exponent 0.43 which implies an asymmetric loss peak fairly close to that predicted here.

\section{General description of viscous liquid dynamics: A proposal}

To simplify matters as much as possible this paper focused on density as the sole relevant configurational variable. In a recent work \cite{dyr05} we discussed the case where the relevant fields are the density and the dipole density fields. This was suggested as a simple model for dielectric relaxation. In that model the density and dipole density fields couple to each other by a third order term in the Hamiltonian (model C of Ref. \cite{hoh77}). In most cases the molecules do not have continuous rotational symmetry around one axis, however, and their orientations should properly be described by more general variables ${\bf\Omega}\in {\rm SO}(3)$ representing the Eulerian angles. The stress tensor is another relevant field which should be included in the description. Finally, the potential energy density should also be included. 

Inspired by the model for density fluctuations discussed in this paper and the above-mentioned model for dielectric relaxation \cite{dyr05}, we  propose the following general recipe for modelling viscous liquid dynamics:

\begin{enumerate}
\item The relevant degrees of freedom are fields $\phi^{(1)}(\br)$, ...,  $\phi^{(n)}(\br)$ defined as: a) the densities of the different types of molecules, b) the densities of the different molecules' orientational variables reflecting their symmetry, c) the six stress tensor fields, and d) the potential energy density.
\item The Hamiltonian $H$ (the free energy) is ultra local; $H$ consists of invariant (i.e., scalar) terms up to some even order.
\item For each field the dynamics are described by a Langevin equation, \[\dot \phi^{(j)}_\bk=-\Gamma^{(j)}_\bk\,\frac{\partial (\beta H)}{\partial \phi^{(j)}_{-\bk}}+\xi^{(j)}_\bk(t)\, ,\] where $\xi^{(j)}_\bk(t)$ is a standard Gaussian white noise term.
\item For each density field the Langevin equation coefficients are given by $\Gamma^{(j)}_\bk=\Gamma^{(j)}_0+D^{(j)} k^2$ where $D^{(j)}\gg\Gamma^{(j)}_0 a^2$, for all remaining fields the rates are k-independent: $\Gamma^{(j)}_\bk=\Gamma^{(j)}_0$.\end{enumerate}

\section{Summary and discussion}

At low temperatures viscous liquid dynamics may be identified with the inherent dynamics taking the system from one potential energy minimum to another. Each such jump is referred to as a flow event. The basic characteristics of a flow event are: 1) Its position; 2) Its time; 3) The induced density change at the position of the flow event quantified by the flow event ``magnitude'' $b$ of Eq. (\ref{5}); 4) The total displacement of all molecule positions induced by the flow event, $\Delta\bf R$. 

The density dynamics discussed in this paper are described by a time-dependent Ginzburg-Landau equation \cite{hoh77} with rates in k space of the form $\Gamma_0+Dk^2$. The first term, which is a consequence of the solidity of viscous liquids, reflects the fact that the density may change locally without changing in the surroundings, a result which implies that density has the appearance of a non-conserved field variable. The second term, $Dk^2$, is the standard diffusion term. It was argued that the disorder of the solid-like liquid and its lack of isotropy most likely result in anisotropic flow events. This is mathematically reflected in the inequality $D\gg\Gamma_0a^2$ which implies an important long-wavelength dominance of the dynamics. Since experimentally there are no long-ranged static density correlations, the long-wavelength dominance of the dynamics makes it realistic to assume that the Hamiltonian is ultra local.

Glarum in 1960 suggested that relaxation takes place via defect motion \cite{gla60,note3}. In his words ``molecules do not relax independently of one another, and the motion of a particular molecule depends to some degree on that of its neighbors... This is because the reorientation of a molecule is far more likely immediately after one of its neighbors has relaxed than it is at an arbitrary time.'' The defect motion was described by a diffusion equation. Glarum's model gave a novel mechanism for explaining dielectric loss peaks which are asymmetric towards the high-frequency side. Inspired by this work Anderson and Ullman in 1967 generally considered the effect of a fluctuating environment on molecular relaxation rates \cite{and67}. These authors showed that fluctuations fast compared to the alpha relaxation rate lead to a symmetric almost Debye response and slow fluctuations to a symmetric, but broad loss peak -- two results which were known already from Kauzmann's 1942 paper on dielectric relaxation \cite{kau42} -- whereas if the environment fluctuates on the alpha time scale, the loss peak becomes asymmetric towards the high-frequency side. Montrose and Litovitz in 1970 \cite{mon70} discussed a model involving an order parameter with both a diffusive and a decaying term in its dynamics. The decay term derives from the fact that ``the structure can change by a simple rate process,'' whereas if for instance the order parameter is a function of the number of holes, the diffusive term corresponds to the ``many small steps arising from the rapid jiggling of molecules.'' In their review from 1972 of the mechanical response of viscous liquid \cite{dex72} Dexter and Matheson summarizes these three papers by the remark: ``Thus, the physical basis of the theories of Glarum, Anderson and Ullman, and Montrose and Litovitz is similar: the molecular environment is assumed to change as a result of spontaneous molecular motion and small diffusional motions.'' The similarity to the present work is clear; here, however, the ``small diffusional motions'' occur in the surroundings of a flow event, and may be thought of as consequences of the ``spontaneous molecular motion.'' After these early works, also in the spirit of the present paper Zwanzig \cite{zwa81} and MacPhail and Kivelson \cite{mackiv} in the 1980's explored the possibility that long-time tail mechanisms are relevant for understanding viscous liquiid dynamics. We mention these works in order to emphasize that the ideas of (III) and this paper have close analogs in several papers published a long time ago.

In this paper we regarded flow events as taking the system instantaneously from one inherent structure to another. Although this approximation applies only on length scales below the solidity length, which is of order 5,000 $\rm\AA$ around the calorimetric glass transition, it is possible that the proposed density dynamics apply also at macroscopic length scales. The problem of properly linking the behavior on length scales below the solidity length to those above needs further consideration, though.

\acknowledgments The author wishes to thank Shankar Das for fruitful discussions on the topic of this paper. This work was supported by the Danish National Research Foundation's grant for the centre for viscous liquid dynamics ``Glass and Time.''

\appendix*\section{}By writing $\langle\rho(\br) \rho(\br')\rangle=\langle\rho\rangle^2+\langle\Delta\rho(\br) \Delta\rho(\br')\rangle$ where $\langle\rho\rangle=N/V$ and $\Delta\rho\equiv\rho-\langle\rho\rangle$, Eq. (\ref{11}) is split into two terms:\begin{equation}\label{A2}\langle|\delta\rk\bk|^2\rangle\,=\,\frac{\langle b^2\rangle}{16\pi^2N}\,\Big[I_1(k)\,+\,I_2(k)\Big]\,,\end{equation}where\begin{equation}\label{A3}I_1(k)\,=\,\int_V\int_V\, \frac{\bk\cdot\br}{r^3}\frac{\bk\cdot\br'}{r'^3}e^{i\bk\cdot(\br-\br')}d\br d\br'\,=\,\left| \int_V\frac{\bk\cdot\br}{r^3}e^{i\bk\cdot\br}d\br\right|^2\,\end{equation}and\begin{equation}\label{A4}I_2(k)\,=\,\int_V\,\frac{\langle\Delta\rho(\br) \Delta\rho(\br')\rangle}{\langle\rho\rangle^2}\frac{\bk\cdot\br}{r^3}\frac{\bk\cdot\br'}{r'^3}e^{i\bk\cdot(\br-\br')}d\br d\br'\,.\end{equation}The latter integral may be expressed in terms of the static structure factor, which for $k\neq 0$ is defined by $S(k)=\langle\rk\bk\romk\rangle=\langle|\rk\bk|^2\rangle$: First note that the definition of $\rk\bk$ (Eq. (\ref{7})) implies that for $k\neq 0$ one has $\sqrt N\rk\bk=\int_V\Delta\rho(\br)\exp(i\bk\cdot\br)d\br$, which implies that $S(k)=1/N\int_V\int_V \langle\Delta\rho(\br) \Delta\rho(\br')\rangle\exp(i\bk\cdot(\br-\br'))d\br d\br'=V/N\int_V \langle\Delta\rho(\bnul) \Delta\rho(\br)\rangle\exp(i\bk\cdot\br)d\br$. Inversion of this Fourier integral allows one to write the density autocorrelation function as a sum over k vectors consistent with periodic boundary conditions: \begin{equation}\label{A5}\frac{\langle\Delta\rho(\br) \Delta\rho(\br')\rangle}{\langle\rho\rangle^2}\,=\,\frac{1}{N}\sum_{\bk'} S(k')e^{i\bk'\cdot(\br-\br')}\,.\end{equation}When this is substituted into Eq. (\ref{A4}) we get\begin{equation}\label{A6}I_2(k)\,=\,\frac{1}{N}\sum_{\bk'} S(k')\left| \int_V \frac{\bk\cdot\br}{r^3}e^{i(\bk+\bk')\cdot\br}\,d\br \right|^2\,.\end{equation}Thus in terms of the integral\begin{equation}\label{A7}I(\bk,\bom)\,\equiv\,\int_V \frac{\bk\cdot\br}{r^3}\,e^{i\bom\cdot\br}\,d\br\,,\end{equation}we have \begin{equation}\label{A7a}\langle|\delta\rk\bk|^2\rangle\,=\,\frac{\langle b^2\rangle}{16\pi^2N}\,\Big[\,\big| I(\bk,\bk)\big|^2\,+\,\frac{1}{N}\sum_{\bk'} S(k') \big| I(\bk,\bk+\bk')\big|^2\,\Big]\,.\end{equation}To evaluate $I(\bk,\bom)$ we note that, if the z-axis is along the $\bom$ vector, $\bk$ is in the xz-plane, and $\thn$ is the angle between $\bom$ and $\bk$, we have $\bom=(0,0,\Omega)$ and $\bk=(k\sin\thn,0,k\cos\thn)$. In spherical coordinates we thus get \begin{equation}\label{A8}I(\bk,\bom)\,=\,\int_0^\pi d\theta\sin\theta\int_0^\infty dr\, r^2\int_0^{2\pi}d\phi\,\frac{kr(\sin\theta\cos\phi\sin\thn+\cos\theta\cos\thn)}{r^3}e^{i\Omega r\cos\theta}\,.\end{equation}This reduces to \begin{equation}\label{A9}I(\bk,\bom)\,=\,2\pi k\cos\thn\,\int_0^\pi d\theta\sin\theta\cos\theta\int_0^\infty dr\, e^{i\Omega r\cos\theta}\,.\end{equation}The radial integral is evaluated by assuming an implicit convergence term, $\lim_{a\rightarrow\infty}\exp(-r/a)$, leading to \begin{equation}\label{A10}I(\bk,\bom)\,=\,2\pi k\cos\thn\,\int_0^\pi d\theta\sin\theta\cos\theta\frac{1}{(-i\Omega\cos\theta)}\,=\, 4\pi i \,\frac{k}{\Omega}\,\cos\thn\,,\end{equation}or\begin{equation}\label{A11}I(\bk,\bom)\,=\, 4\pi i \, \frac{\bk\cdot\bom}{\Omega^2}\,.\end{equation}In conclusion, Eq. (\ref{A7a}) becomes\begin{equation}\label{A12}\langle|\delta\rk\bk|^2\rangle\,=\,\frac{\langle b^2\rangle}{N}\,\Big[1\,+\,\frac{1}{N}\sum_{\bk'} S(k') \left(\frac{\bk\cdot(\bk+\bk')}{(\bk+\bk')^2}\right)^2\,\Big]\,.\end{equation}


\begin{thebibliography}{99}
\bibitem{note1} One usually defines a solid as a system which does not flow, implying that a ``solid that flows'' is a contradiction in terms. Whether a system flows or not, however, is a matter of time scale; even a crystal flows at finite temperatures: In a gravitational field, for instance, a horizontally flat macroscopic crystalline sample has lower free energy than, e.g., a cubic-shaped sample. Since any crystal in thermal equilibrium contains point defects, and since their movements are impeded only by finite energy barriers, ``flow'' from the cubic-shaped sample to the flat is not only possible, but bound to take place over sufficiently long time scales.
\bibitem{rho_note} Note that with this rate Eq. (\ref{13}) in real space has the following form: $\dot\rho(\br)\propto(-\Gamma_0+D\nabla^2)\delta(\beta H)/\delta\rho(\br)+{\rm white\, noise}$, an equation which does not obviously make sense for an ultra local Hamiltonian, but which makes perfect sense when the theory is put on a lattice.
\end{thebibliography}
\end{document}